\documentclass[aps,twocolumn,showpacs,superscriptaddress,floatfix]{revtex4-1}
\usepackage{graphicx,amsfonts,amssymb,amsmath,mathrsfs,hyperref,bm}

\usepackage{siunitx}
\usepackage{epstopdf}
\usepackage{xcolor}
\begin{document}
\title{Quantum nonlinear planar Hall effect in bilayer graphene: an orbital effect of a steady in--plane magnetic field}

\author{N. Kheirabadi}
\affiliation{Department of Physics, Sharif University of Technology, P.O.Box 11155-9161, Tehran, Iran}
\email{kheirabadinarjes@gmail.com}

\author{A. Langari}
\affiliation{Department of Physics, Sharif University of Technology, P.O.Box 
11155-9161, Tehran, Iran}
\email{langari@sharif.edu}


\begin{abstract}
We study the quantum nonlinear planar Hall effect in bilayer graphene under a steady in--plane magnetic field. When time--reversal symmetry is broken by the magnetic field, a charge current occurs in the second--order response to an external electric field, as a result of the Berry curvature dipole in momentum space. We have shown that a nonlinear planar Hall effect originating from the anomalous velocity is deduced by an orbital effect of an in--plane magnetic field on electrons in bilayer graphene in the complete absence of spin--orbit coupling. 
Taking into account the symmetry analysis, we derived the dominant dependence of Berry curvature dipole moment on the magnetic field components. 
Moreover, we illustrate how to control and modulate the Berry curvature dipole with an external planar magnetic field, gate voltage, and Fermi energy. 
\end{abstract}

\maketitle
\section{Introduction}\label{Introduction}

During the last century, the Hall effect has played an important role in the advance of technology 
and condensed matter physics \cite{ho2021hall} and because of their profound relation to the topology, the Hall effects family have been diligently scrutinized in the recent years \cite{landau2013statistical, xiao2010berry, moore2010confinement, sodemann2015quantum}. 

When an electric field drives a current through a crystal, the system is out of equilibrium and the electron velocity originates from the group velocity of the electron wave packet, while the anomalous velocity arises from the Berry curvature, which is an intrinsic property emerging from the band structure. 
The conventional Hall conductivity, the quantization of the Hall conductance in strong magnetic fields, can be considered as the zero order moment of the Berry curvature over occupied states \cite{sodemann2015quantum}.
The linear anomalous Hall effect and quantum anomalous Hall effect have recently been observed in 
topological materials  with broken time--reversal symmetry, such as, magnetically doped topological insulators \cite{liu2008quantum, yu2010quantized, chang2013experimental} and magnetic Weyl semimetals \cite{xu2011chern, burkov2014chiral, zhang2018electrically}.

The first--order moment of the Berry curvature over the occupied states is defined by the Berry curvature dipole (BCD), which is a pseudo tensor leads to the quantum nonlinear Hall effect
 \cite{sodemann2015quantum}.
It has been shown that up to the second order, and dissimilar to the linear effects, the quantum nonlinear Hall effect shows a component of the voltage oscillating at twice the frequency of the driving alternating electric field (the second--harmonic Hall voltage) and a steady component that is caused due to the rectification effect, by which an AC electric field is turned into a DC signal \cite{sodemann2015quantum}. 
The quantum nonlinear Hall effect 
has been distinguished in 1Td WTe2  \cite{ma2019observation, kang2019nonlinear, ho2021hall} and has been predicted to happen in some developing materials with low crystalline symmetries  \cite{sodemann2015quantum, zhang2018electrically, shi2019symmetry}. 
For the two-dimensional crystals with trigonal symmetry in the presence of in-plane magnetic field, a nonzero BCD also leads to a topological response in the nonlinear planar Hall effect \cite{battilomo2021anomalous}.  

In the planar Hall effect (PHE), in contrast to the ordinary Hall effect, the transverse voltage arises when an in--plane magnetic field is applied. In this regime, the applied electric field, the magnetic field, and the transverse Hall voltage  are in the same plane, contrary to the arrangement in which the conventional Hall effect vanishes. In the most 2--dimensional (2D) materials, the PHE has an absolutely  semiclassical origin. 
In thin films of antiferromagnetic semiconductor,  the observed PHE is suggested to be the result of band anisotropies \cite{yin2019planar}. 
It has also been indicated that PHE appears in 2D-electron gases on the interfaces of perovskite oxides \cite{wadehra2020planar, joshua2013gate} and thin films of ferromagnetic semiconductors \cite{tang2003giant, bowen2005order, ge2007magnetization}.
Moreover, the PHE perform an important role in the transport properties of Weyl semimetals. 
Recently, it has been shown that the Zeeman--induced nontrivial Berry curvature affects the PHE in 2D trigonal crystals \cite{battilomo2021anomalous}. However, the common aspect of all of these cases is that the PHE arises form magnetic materials or spin-orbit origins.

In time--reversal invariant materials, BCD is the effect of spin--orbit coupling or warping of the Fermi surface \cite{ortix2021nonlinear}. 
The quantum nonlinear planar Hall effect (QNLPHE) we discuss here is determined by BCD when time--reversal is broken by an applied in--plane magnetic field. This  QNLPHE has a quantum effect arising from the anomalous velocity of Bloch electrons generated by the Berry curvature, which is not quantized \cite{sodemann2015quantum}. In this study, we show for the first time that a non--zero BCD is achievable in the complete absence of spin effects in bilayer graphene; an excellent 2D material candidate with giant intrinsic carrier mobilities and a tunable band gap \cite{morozov2008giant, ohta2006controlling}. Besides, our results represent a 
distinct theoretical demonstration of a BCD, which can be manipulated by magnetic fields. Such a tunable BCD deduces to a broad range of quantum geometrical phenomena such as the magnetically switchable circular photogalvanic effect \cite{xu2018electrically}, and rectification \cite{ideue2017bulk}.
For all of the mentioned cases, a nonzero BCD is a requirement, which makes them fundamentally important and interesting  \cite{xu2018electrically}. In this article,  we calculate numerically the BCD of bilayer graphene imposed by an in-plane magnetic field and obtain an expression, which shows the dependence on the components of magnetic field as far as space inversion is fulfilled. 

The structure of this article is as follows.
In the next section, we review the basic notions of BCD. In Sec.~\ref{HS} we introduce the Hamiltonian of bilayer graphene with an in-plane magnetic field. Although the magnetic field breaks time reversal symmetry, for zero gate voltage the space inversion symmetry is valid, which leads to an expression for BCD components with respect to 
the components of magnetic field.  This expression shows the dominant behaviour of magnetic field. We present the numerical results of our model in Sec.~\ref{simulations}, where we investigate the dependence of BCD on the energy gap, Fermi energy and magnetic fields, which justifies our analytic findings. We discuss some practical aspects of our results in Sec.~\ref{discussion} and conclude our finding is Sec.~\ref{conclusion}.

\section{Berry curvature dipole moment}\label{bcd}
For an applied in--plane oscillating electric field with angular frequency $\omega$,
$\vec{E}(t)=Re\{\vec{E} e^{i \omega t}\}$. Based on the Boltzmann transport approach, and as we mentioned in Section.~\ref{Introduction}, it has been shown that two currents in second order of electric field that originate from the anomalous velocity of electrons are anticipated in a crystal; $j_a=Re\{j_a^0+j_a^{2 \omega} e^{2 i \omega t}\}$ \cite{sodemann2015quantum}. Here, $j_a^0$ is the DC response to the applied oscillating electric field and $j_a^{2 \omega}$ is the second harmonic generation current. For 2D materials, like bilayer graphene, it has been shown that the AC and DC currents have the following forms \cite{battilomo2021anomalous, sodemann2015quantum}
\begin{eqnarray}
\vec{j}^0&&=\frac{e^3 \tau}{2 \hbar^2(1+i \omega \tau )}\hat{z} \times \vec{E}^* \big( \vec{D} \cdot \vec{E}\big),\nonumber\\
\vec{j}^{2 \omega}&&=\frac{e^3 \tau}{2 \hbar^2(1+i \omega \tau )}\hat{z} \times \vec{E} \big( \vec{D} \cdot \vec{E}\big).
\end{eqnarray} 
In the above equations, $e$ is the absolute value of electron charge, $e>0$, $\tau$ is the scattering time and $\mathbf{D}$ is the dipole moment of the Berry curvature over the occupied states, BCD, which is equal to 
\begin{eqnarray}\label{bcd}
D_{a}= \int_k{f_0(\partial_a \Omega_z)},
\end{eqnarray}
where $\int_k\equiv\int d^2k/(2 \pi)^2$, and $f_0$ is the equilibrium Fermi--Dirac distribution function \cite{sodemann2015quantum}. At zero temperature, $f_0=\Theta(\mu-\epsilon(\mathbf{k}))$ where $\epsilon$ is the energy dispersion of electrons and $\mu$ is the Fermi level; so, $f_0=1$ if $\epsilon(\mathbf{k})<\mu$, else $f_0=0$. 
In 2D materials, the Berry curvature is a pseudoscalar that only have an out-of-plane component. 
Consequently, BCD is a pseudovector confined in the corresponding 2D plane and is normalized to unit length
\cite{sodemann2015quantum}. 
Furthermore, for 2D materials, the Berry curvature of the $n$-th band is defined by the following equation
\begin{eqnarray}\label{omegan}
\Omega_z^n(\mathbf{k})= i \sum_{n \prime \neq n}\frac{\langle n|\partial_x \hat{H}|n^\prime\rangle\langle n^\prime|\partial_y \hat{H}|n\rangle-(x\leftrightarrow y)}{(\epsilon_n-\epsilon_{n^\prime}) }
\end{eqnarray}
where $\epsilon _n$ is the eigenvalue of the Hamiltonian and $\partial_{x/y}\equiv\partial_{k_x/k_y}$. 
It is necessary to obtain the eigenstates and eigenvalues of the Hamiltonian of bilayer graphene in a parallel magnetic field
to derive the Berry curvature and its corresponding dipole.
In the next section, we introduce the Hamiltonian of bilayer graphene in an in--plane steady magnetic field and we consider the effect of the symmetry on the general form of the BCD in bilayer graphene.
However, the symmetry analysis, which appears in the next section is also valid for any 2D material with broken time-reversal symmetry while space inversion symmetry is satisfied. 
\section{Hamiltonian and Symmetry Analysis}\label{HS}
\begin{figure}
   \centering   
   \includegraphics[scale=0.45]{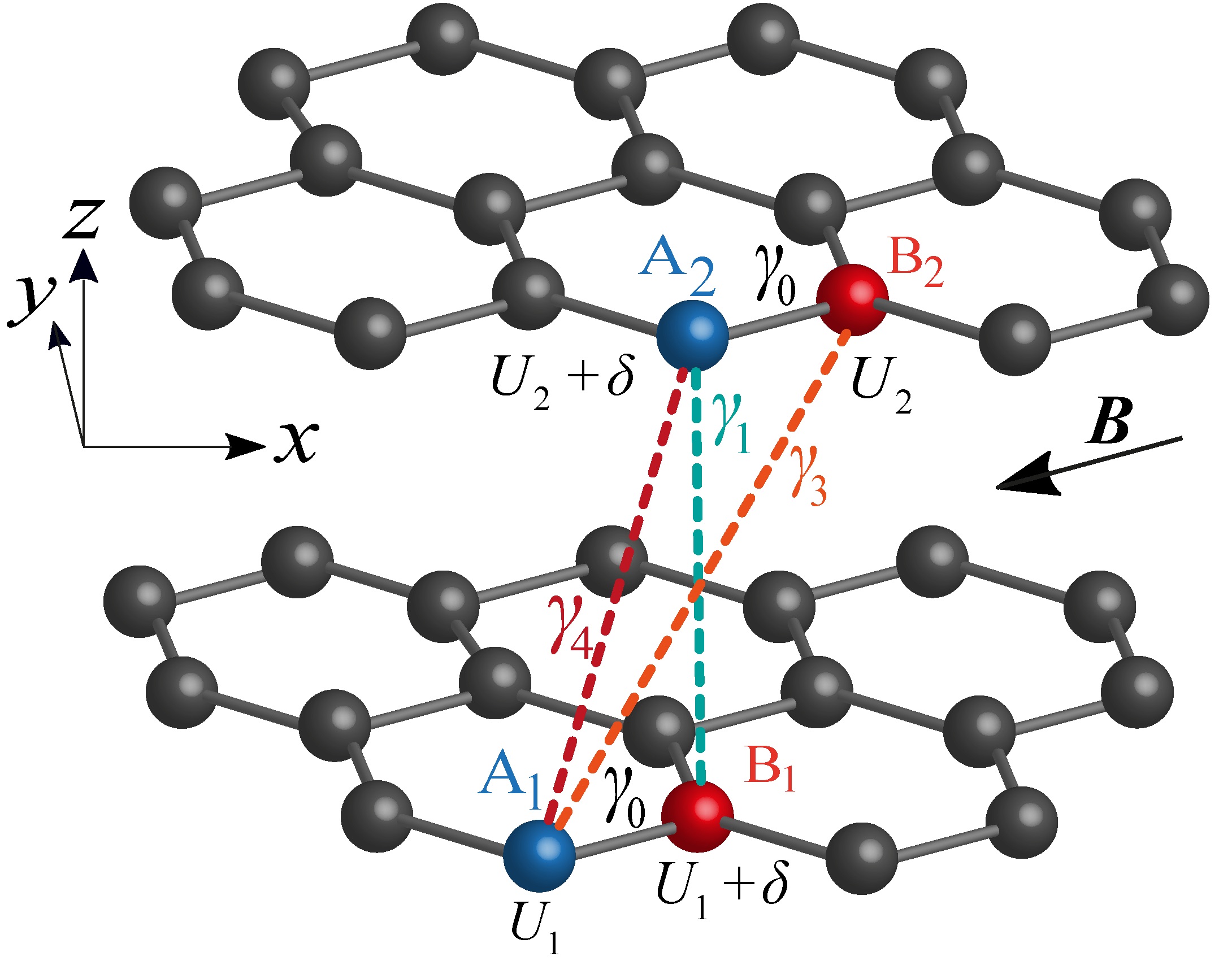}
   \caption{The AB--stacked bilayer graphene unit cell. $A_1$ and $B_1$ atoms on the bottom layer and $A_2$ and $B_2$  on the top layer have been depicted. Straight lines point out intralayer coupling $\gamma_0$, vertical dashed lines also show interlayer coupling $\gamma_1$ and skew interlayer couplings $\gamma_3$ and $\gamma_4$. Parameters $U_1$, $U_2$, and $\delta$ indicate different on--site energies.} 
    \label{ab}
\end{figure}

\begin{figure}
   \centering   
   \includegraphics[scale=0.50]{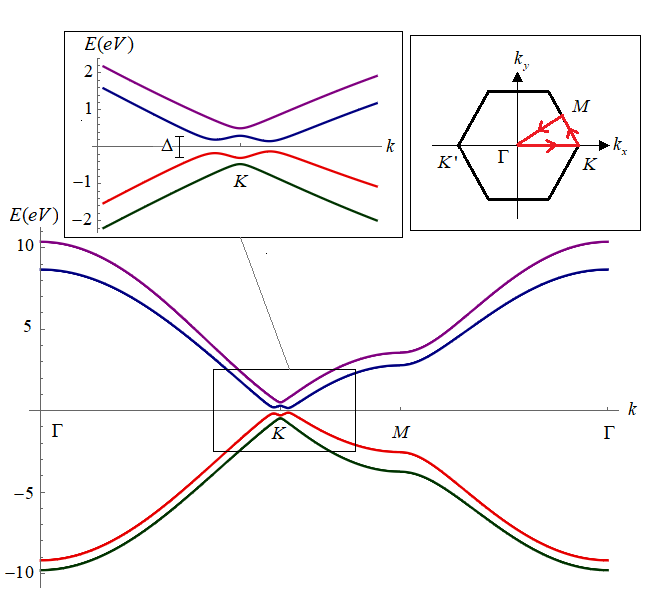}
   \caption{The band structure of bilayer graphene along $\Gamma$ - $K$ - $M$ - $\Gamma$ direction denoted by the red line in the right inset within the first Brillouin zone of the bilayer graphene. The left inset indicates the opening of gap near $K$ point; $\Delta=U_2-U_1= 0.6$ eV. The parameters used to calculate the band structure are $\gamma_0 = 3.16$ eV, $\gamma_1 = 0.389$ eV, $\gamma_3=0.384$ eV, $\gamma_4= 0.14$ eV, $\delta = 0.018$ eV, $a=2.46$ \si{\angstrom}, and $d \approx 3.3$
\si{\angstrom} \cite{kuzmenko2009determination, mccann2013electronic}.}
    \label{band}
\end{figure}

The AB--stacked bilayer graphene structure and its related parameters with a lattice constant that is equal to $a$ and interlayer distance $d$ is depicted in the Fig.~\ref{ab}. According to this figure, different on--site
energies are $U_1$ and $U_2$ on the $A_1$ and $B_2$ sites, respectively, which are the on--site energies of the two layers. $\delta$ is also an energy difference between $A$ and
$B$ sublattices on each layer. Based on the tight--binding approximation, we have considered the full Hamiltonian of bilayer graphene considering hopping parameters $\gamma_0, \gamma_1, \gamma_3, \gamma_4$  and on--site energies stated in Fig.~\ref{ab}. The band structure of bilayer graphene for a specific set of parameters is depicted in Fig.~\ref{band}.  The difference of layers bias, $\Delta=U_2-U_1$ generates a finite gap
at the $K$-point of the first Brillouin zone.

To derive the Hamiltonian of bilayer graphene in a parallel magnetic field, the in-plane magnetic field has been considered as a phase change given by a path integral of $\mathbf{A}$, vector potential of the in--plane magnetic field. For, example, to calculate $H_{AB}$ elements of the Hamiltonian, related to the $\gamma_0$ hopping parameter in the tight-binding approach, and by considering nearest--neighbour approximation, we have assumed that each A atom has three nearest-neighbour B atoms. Consequently, $H_{AB}$ is
\begin{eqnarray}
H_{AB}=-\gamma_0 \sum_{j=1}^3 {\exp \bigg( i \bf{k}. (\bf{R}_{Bj}-\bf{R}_A)-\frac{ie}{\hbar}\int_{\bf{R}_{Bj}}^{\bf{R}_A} \bf{A}.\bf{d}l \bigg)}.\nonumber\\
\end{eqnarray}
where $\mathbf{A}$ has been chosen to be $z(B_y,-B_x)$ to keep translation symmetry in the grapehen plane. Details of the derivation to reach the final Hamiltonian can be found in 
Ref. \cite{kheirabadi2016magnetic}.

Hence, the $4 \times 4$ Hamiltonian of bilayer graphene in a steady parallel magnetic field in the basis of $(A_1,B_1,A_2,B_2)^T$ is \cite{kheirabadi2018electronic, kheirabadi2016magnetic}
\begin{eqnarray}\label{Hamiltonian}
H=\begin{pmatrix}
U_1 &-\gamma_0 f_1(\textbf{k}) &\gamma_4 f(\textbf{k}) &-\gamma_3f^*(\textbf{k}) \\ 
-\gamma_0 f_1^*(\textbf{k}) &U_1+\delta &\gamma_1 &\gamma_4 f(\textbf{k}) \\ 
\gamma_4 f^*(\textbf{k}) &\gamma_1  &U_2+\delta &-\gamma_0 f_2(\textbf{k}) \\ 
-\gamma_3f^(\textbf{k}) &\gamma_4 f^*(\textbf{k}) &-\gamma_0 f_2^*(\textbf{k})  & U_2
\end{pmatrix},\nonumber\\
\end{eqnarray}
where 
\begin{eqnarray}
f&&=\exp\big(\frac{i k_y a}{\sqrt{3}}\big)+2 \exp\big(-\frac{i k_y a}{2 \sqrt{3}}\big) \cos \big(\frac{k_x a}{2}\big),\\
f_1&&=\exp\big(\frac{i k_y a}{\sqrt{3}}+i e \frac{a d }{2 \sqrt{3} \hbar} B_x\big)\nonumber\\
&& + \exp\big( -i \big( -\frac{k_x a}{2}+\frac{k_y a}{2 \sqrt{3}}\big)-i e \frac{a d }{2 \hbar} \big( \frac{B_y}{2}+\frac{B_x}{2 \sqrt{3}}\big)\big)\nonumber\\
&& + \exp\big( -i \big( \frac{k_x a}{2}+\frac{k_y a}{2 \sqrt{3}}\big)+i e \frac{a d }{2 \hbar} \big( \frac{ B_y}{2}-\frac{B_x}{2 \sqrt{3}}\big)\big),\\
f_2&&=\exp\big(\frac{i k_y a}{\sqrt{3}}-i e \frac{a d }{2 \sqrt{3} \hbar} B_x\big)\nonumber\\
&& + \exp\big( -i \big( -\frac{k_x a}{2}+\frac{k_y a}{2 \sqrt{3}}\big)+i e \frac{a d }{2 \hbar} \big(\frac{ B_y}{2}+\frac{B_x}{2 \sqrt{3}}\big)\big)\nonumber\\
&& + \exp\big( -i \big( \frac{k_x a}{2}+\frac{k_y a}{2 \sqrt{3}}\big)-i e \frac{a d }{2 \hbar} \big( \frac{ B_y}{2}-\frac{B_x}{2 \sqrt{3}}\big)\big).
\end{eqnarray}
Here, $\textbf{k}$ is the electron wave vector, and $\textbf{B}$ is the magnetic field vector, i.e.  $\mathbf{B}=\left(B_x,B_y,0\right)$. We assume that the lower layer of the bilayer is located at $z=-d/2$ and the upper layer is stated at $z=+d/2$.

The time--reversal symmetry of the Hamiltonian,  Eq.~\ref{Hamiltonian}, is broken by the planar magnetic field, because apparently $H^*(\mathbf{k}) \neq H(-\mathbf{k})$. On the other hand, $H(\mathbf{k})$ satisfies spatial inversion if  
$\mathcal{U} H (\mathbf{k}) \mathcal{U}^\dagger = H(-\mathbf{k})$; where, $\mathcal{U}$ operator swap $A_1 \leftrightarrow B_2$ and $B_1 \leftrightarrow A_2$. 
It can be shown that for $U_1=U_2$ or in the absence of a gate voltage, the Hamiltonian is invariant under spatial inversion. 

In order to resolve the symmetry properties of BCD in bilayer graphene imposed by in plane magnetic field,
we consider the following expansion for BC in terms of the magnetic filed components,
\begin{eqnarray}\label{BC}
\Omega(k_x,k_y,B_x,B_y)=\sum_{m, n \geqslant 0}{a_{m, n}(k_x,k_y) {B_x}^m {B_y}^n},
\end{eqnarray}  
where $m, n$ are integers, $a_{m, n}(k_x,k_y)$ are coefficients of the expansion, which are functions of momentum ($k_x, k_y$). 
Since Eq.(\ref{BC}) represents BC in the presence of a magnetic field $a_{0, 0}(k_x, k_y)=0$.
The Berry curvature is invariant under inversion symmetry ($\mathcal{U}$, which can be recognized as $x \rightarrow -x$ and $y \rightarrow -y$) that leads to the following constraint for the $a_{m, n}$ coefficients:
\begin{equation}
a_{m, n}(k_x,k_y)= (-1)^{m+n} a_{m, n}(-k_x,-k_y).
\label{amn_inv}
\end{equation}
Moreover, taking into account the reflection symmetry with respect to $y$-axis ($x \rightarrow -x$) or
$x$-axis ($y \rightarrow -y$) gives the following relations, respectively
\begin{eqnarray}
a_{m, n}(k_x,k_y)&=& (-1)^n a_{m, n}(-k_x,k_y), \nonumber \\
a_{m, n}(k_x,k_y)&=& (-1)^m a_{m, n}(k_x,-k_y).
\label{amn_ref}
\end{eqnarray}
According to Eq.(\ref{bcd}),  the BCD is obtained by integrating the derivatives of BC over the Brillouin zone.
Making use of Eqs.(\ref{amn_inv}, \ref{amn_ref}), we show that the terms in BCD expansion, which contain 
both even or odd values of $m$ and $n$ vanish. Hence, one of the exponents (either $m$ or $n$) must be odd.
The proof of this statement is presented in Appendix.~\ref{A}.
Accordingly, the BCD of our model in the presence of an in-plane magnetic field has the 
following form:
%
%
%
\begin{eqnarray}\label{dx}
D_x= B_y \int_k f_0 \sum_{m, n \geqslant 0}\frac{\partial a_{2m,2n+1}(k_x, k_y)}{\partial k_x} {B_x}^{2m}{B_y}^{2n},
\end{eqnarray}
\begin{eqnarray}\label{dy}
D_y= B_x \int_k f_0 \sum_{m, n \geqslant 0}\frac{\partial a_{2m+1,2n}(k_x, k_y)}{\partial k_y} {B_x}^{2m}{B_y}^{2n}.
\end{eqnarray}
Hence, whenever the inversion symmetry is satisfied in bilayer graphene, for a magnetic field in the $x (y)$ direction a non--zero $D_y (D_x)$ is predicted. It means that for a non--zero $B_x (B_y)$ at $B_y (B_x)=0$ T, 
$D_y (D_x)$ should be the only non--zero component of BCD, which shows a linear dependence on $B_x (B_y)$.
This argument is also valid for any 2D material with broken time-reversal symmetry while space inversion symmetry is satisfied.
\section{Tunable BCD in bilayer graphene}\label{simulations}
In this section, we present numerical results of BCD for bilayer graphene in the presence of an in-plane magnetic field.
We use the following parameter values in our numerical calculations: $\gamma_0 = 3.16$ eV, $\gamma_1 = 0.389$ eV, $\gamma_3=0.384$ eV, $\gamma_4= 0.14$ eV, and $\delta = 0.018$ eV, and the lattice spacing is $a=2.46$ \si{\angstrom} and interlayer spacing is $d \approx 3.3$
\si{\angstrom} \cite{kuzmenko2009determination, mccann2013electronic}.
Our numerical results verify the the general form of BCD presented in Eqs.(\ref{dx}, \ref{dy}).
In other words,  as the magnetic field approaches zero value the BCD of our system vanishes regardless of the position of the Fermi energy or any applied gate voltage; so, the deduced Hall effect is a genuine Hall effect. Moreover, a linear dependence of BCD on either $B_x$ or $B_y$ is also observed.
The detail of the numerical calculations is described in Appendix.~\ref{BB}. 

\begin{figure}
   \centering   
   \includegraphics[scale=0.5]{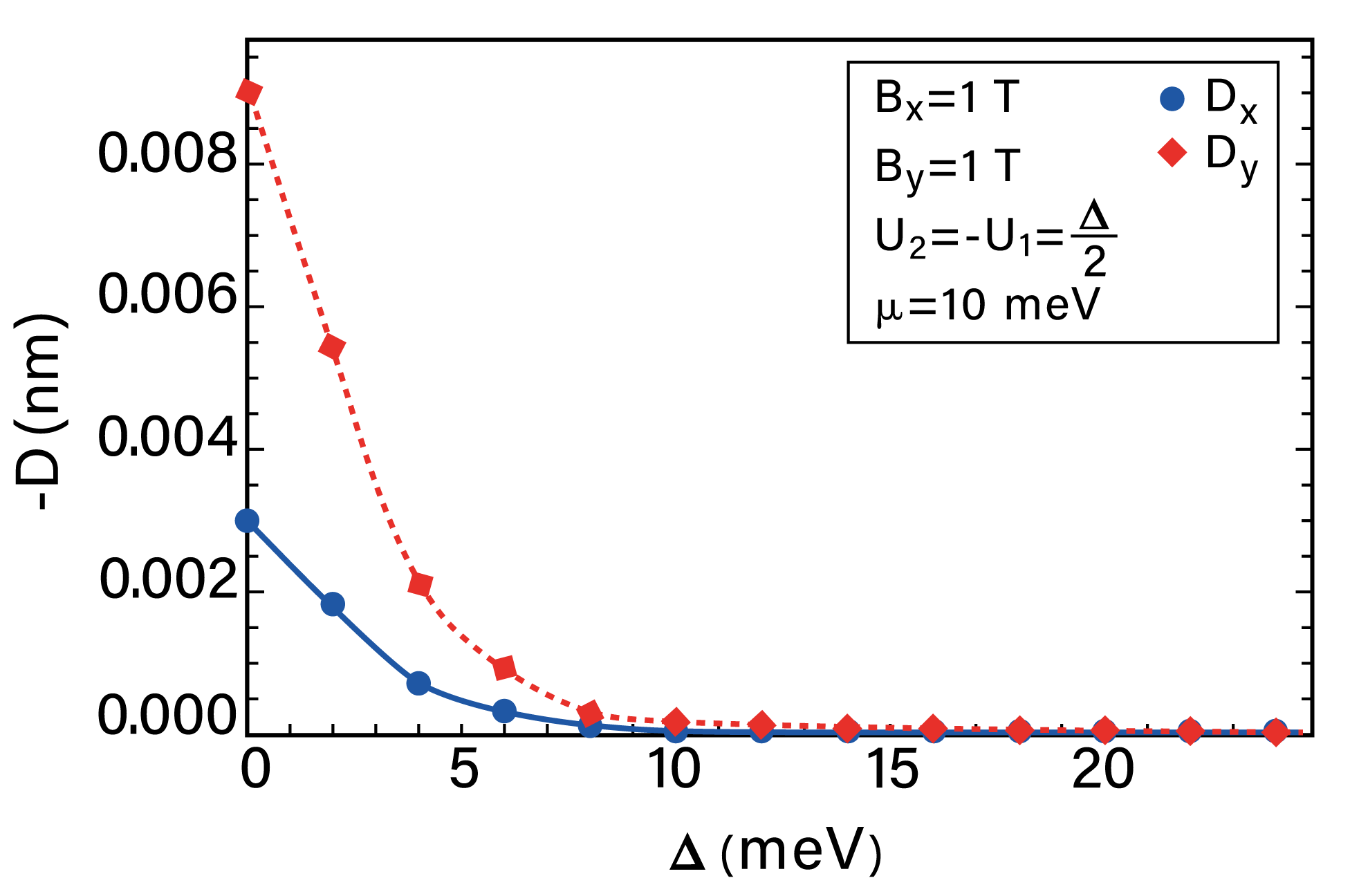}
   \caption{The change of $D_x$ ($D_y$) versus $\Delta$ (the on--site energies $U_2=-U_1=\Delta/2$), depicted by a solid line (dashed line) where $B_x=B_y = 1$ T. The Fermi energy is $10$ meV.} \label{Delta}
\end{figure}

\subsection{Gap dependent BCD}
Early studies on the TaAs-family of Weyl semimetals has shown that a zero or a small gap region in the band structure leads to a large BCD \cite{PhysRevB.97.041101}. The relation between the gap and BCD could be understood based on Eq.~\ref{omegan}, where the smaller gap in the denominator causes the larger value of BCD. Accordingly, the control on the band structure and wave functions will come out with the control on BCD \cite{zhang2018electrically}.
For bilayer graphene in a planar steady magnetic field, an applied in--plane magnetic field opens a gap in the band structure of bilayer graphene. For example, in the absence of any applied external gate voltage; $U_1=U_2=0$ eV and at $B_x=B_y=1$ T, a gap of the order of $10^{-5}$ meV is deduced in our system. Although the amount of gap is small it leads to a large gradient of the Berry curvature of bilayer graphene. 

We have plotted in Fig.~\ref{Delta}, the $x-$ and $y-$component of BCD of bilayer graphene for $B_x=B_y=1$ T versus the on-site energy  $\Delta$ defined by $U_2=- U_1=\frac{\Delta}{2}$ at the chemical potential $\mu=10 $ meV.
Both components of BCD show monotonically decreasing behaviour versus $\Delta$, where the maximum is 
at $\Delta=0$, which justifies the effect of magnetic field to produce BCD.
By applying a gate voltage (presented by the on-site energies), the gap of system is dominated by the effect of $\Delta$ rather than the gap created by the magnetic field. Increasing $\Delta$ washes out the BCD created by magnetic field, which is clearly seen for $\Delta > 10$ meV in Fig.~\ref{Delta}.

\subsection{Fermi energy dependence}

\begin{figure}
   \centering   
   \includegraphics[scale=0.5]{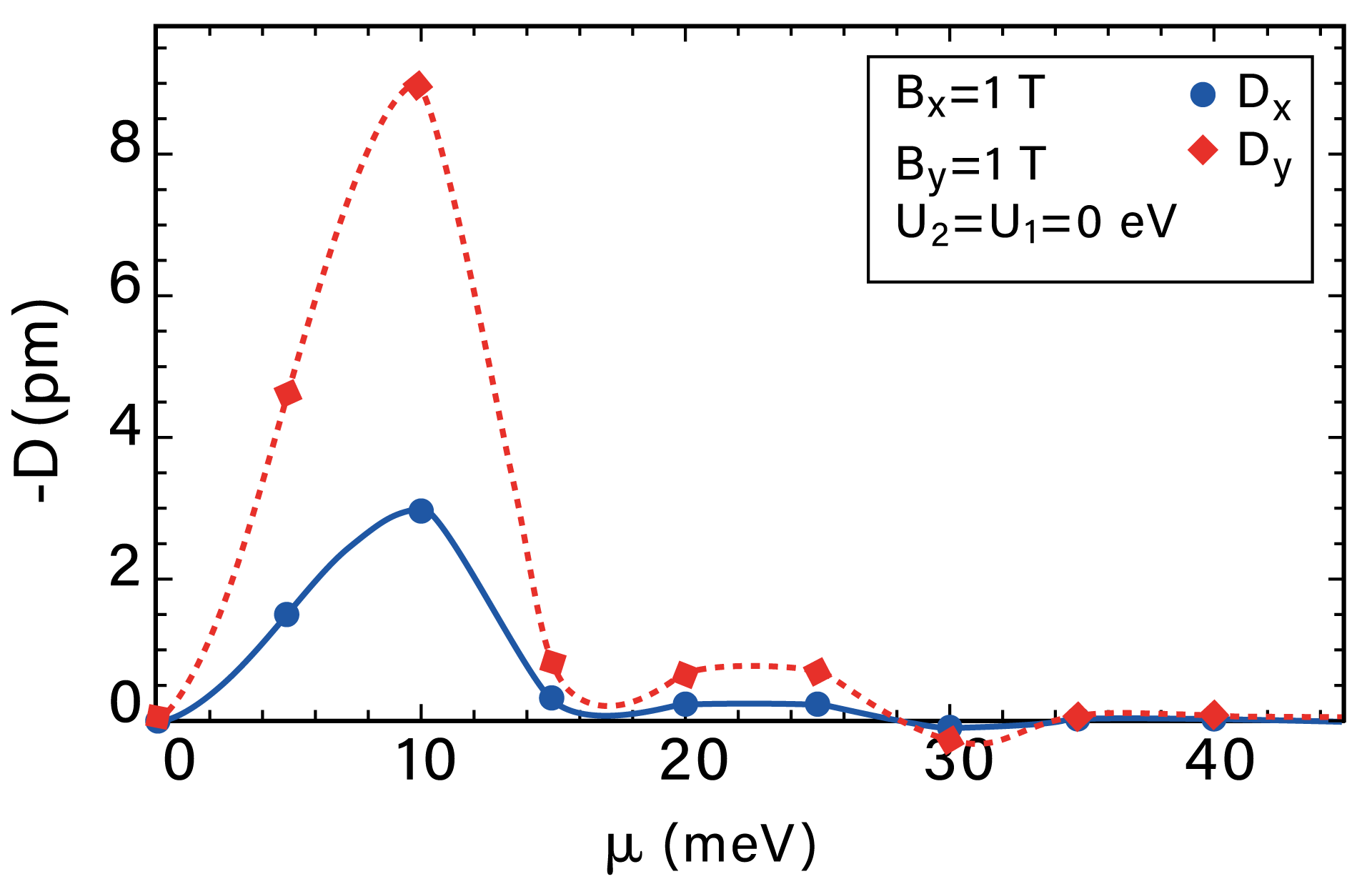}
   \caption{The variation of BCD versus the Fermi level, $\mu$. The solid (dashed) line depicts $D_x$ ($D_y$) where $(B_x, B_y) = (0, 1)$ T ($(B_x, B_y)=(1, 0)$ T) and $U_2=U_1=0$ eV. Whenever, both components of magnetic field is non--zero, $(B_x, B_y)=(1, 1)$ T, both solid and dashed plots should be considered for the $D_x$ and $D_y$, respectively.} 
    \label{Mu}
\end{figure}

The Berry curvature dipole is a Fermi surface property, which depends on the position of the Fermi energy. This is the motivation to obtain the BCD for different values of chemical potential. 
We have plotted in Fig.~\ref{Mu} both $D_x$ and $D_y$ versus the Fermi energy ($\mu$) for $(B_x, B_y)= (1, 1)$ T
and zero on-site bias $U_1=U_2=0$. Both plots show non-monotonic behaviour versus $\mu$, where the maximum BCD
appears at $\mu=10$ meV. The asymmetry between $D_x$ and $D_y$ for equal components of magnetic field is due
to the asymmetry of geometry of lattice, where the x-direction is along the zigzag edges.  BCD is zero at $\mu=0$, where the Fermi energy is in the middle of gap between filled and empty bands.  According to the expression of BCD (Eq.~\ref{bcd}), for the latter case we would obtain zero. This means that only partially filled band(s) contribute to a
non--zero BCD.

Moreover, we have also calculated BCD for $(B_x, B_y)= (1, 0)$ T, and $U_2-U_1=\Delta=0$ eV. In agreement with our results in Sec.~\ref{HS}, a non--zero BCD is observed only in $y$-direction, which falls on the dashed line in Fig.~\ref{Mu}, while $D_x$ is zero. If we switch the components of the magnetic fields to $(B_x, B_y)= (0, 1)$ T, a non--zero BCD in $x$-direction is deduced that is parallel to zigzag edges (solid line in Fig.~\ref{Mu}), while $D_y=0$.
The symmetry analysis of Sec.~\ref{HS}, which led to Eqs.~(\ref{dx}, \ref{dy}) is justified by our numerical results presented in Fig.~\ref{Mu}. In addition, the overlap of $D_x$ in Fig.~\ref{Mu} for $(B_x, B_y)= (1, 1)$ T with 
the case of $(B_x, B_y)= (0, 1)$ T means that in 
the general form of BCD (Eqs.~(\ref{dx}, \ref{dy}))  the linear term of expansion has the dominant effect.
This is also the case of $D_y$ for $(B_x, B_y)= (1, 1)$ and  $(B_x, B_y)= (1, 0)$.

\subsection{BCD dependence on the in-plane magnetic field}\label{B-dependence}

\begin{figure}
   \centering   
   \includegraphics[scale=0.15]{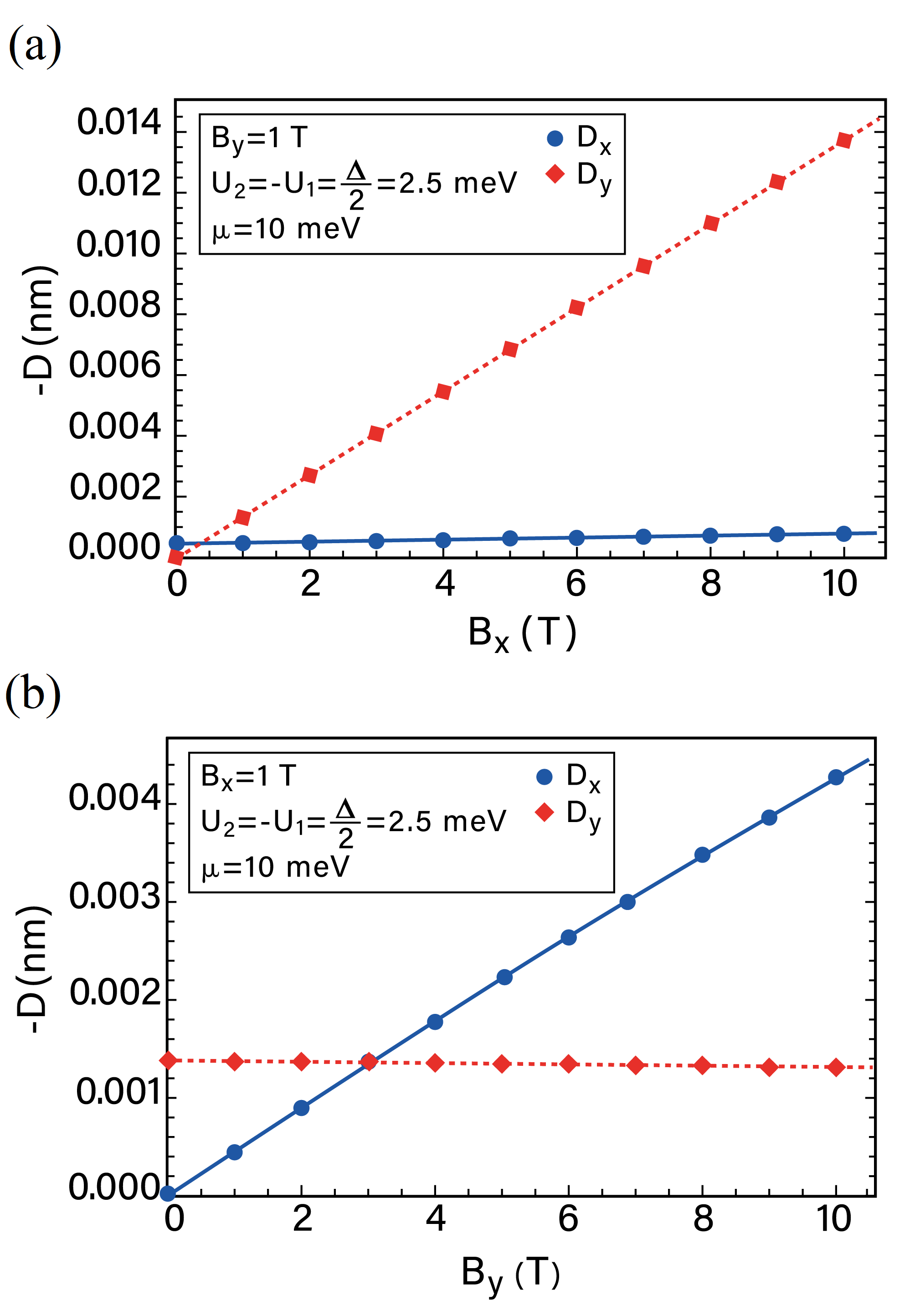}
   \caption{The variation of $D_x$ ($D_y$) versus the in-plane magnetic field are shown by a solid line (dashed line), where the Fermi energy is $10$ meV and $U_2=-U_1=\Delta/2=2.5$ meV. (a) The magnetic field in $y$-direction is constant, $B_y=1$ T,  while $B_x$ changes from $0$ to $10$ T. (b) The magnetic field in $x$-direction is fixed to $B_x=1$ T, while the other component $B_y$ changes form $0$ to $10$ T.} 
    \label{B}
\end{figure}

Here, we study the effect of a steady in-plane magnetic field on BCD for a non-zero gate voltage.
It has to be mentioned that a non-zero gate potential ($\Delta\neq0$) breaks the inversion
symmetry of our model and we can not rely on the arguments led to Eqs.~\ref{dx} and \ref{dy}.
Accordingly, we considered two cases at non-zero on-site energy $\Delta=5$ meV, (a) the effect of $B_x$ on BCD for fixed $B_y=1$ T, and (b) the response to $B_y$ at fixed $B_x=1$ T.
 
We have plotted both $D_x$ and $D_y$ versus $B_x$ at fixed $B_y=1$ T and $\Delta=5$ meV, in Fig.~\ref{B}-(a). We observe a non-zero and almost a constant value for $D_x$ even at $B_x=0$, which shows that it is mainly controlled by the fixed value of $B_y$. However, $D_y$ shows a linear behaviour versus $B_x$, which resembles the leading term obtained in Eq.~(\ref{dy}).

In Fig.~\ref{B}-(b), the components of BCD has been plotted versus $B_y$ at constant values of $B_x=1$ T and $\Delta=5$ meV. Similar to the case (a), $D_y$ is non-zero and constant versus $B_y$, proposing its dependence on $B_x$. Moreover, $D_x$ has an almost linear dependence on $B_y$, which looks like the leading term of Eq.~\ref{dx}. It has to be mentioned that the zigzag direction of graphene is along the $x$-axis, which breaks the symmetry by exchanging $x \leftrightarrow y$.

Although a non-zero gate voltage breaks the space inversion symmetry of our model the numerical results reveal that the dependence of $D_x (D_y)$ on magnetic fields is dominated by the leading terms given in Eq.~\ref{dx} (Eq.~\ref{dy}).

\section{Discussion}\label{discussion}

An applied in-plane magnetic field has two aspects on bilayer graphene: (i) it breaks the time--reversal symmetry, which (ii) opens a band gap although being small leads to a large BCD. A symmetry analysis based
on the spatial inversion symmetry concludes to BCD dependence on the components of an in-plane magnetic field, which has been presented in Eqs.~(\ref{dx}, \ref{dy}). Although a non-zero gate voltage (on-site energies) breaks the spatial inversion symmetry, our numerical results in Sec.~\ref{B-dependence}
render that the magnetic field dependence of BCD is dominated by the first term in Eqs.~(\ref{dx}, \ref{dy}).

We have mentioned in Section.~\ref{bcd} that a non--zero BCD deduces to two types of the second order currents; a DC one and an AC one. The magnitude of these terms are proportional to the second order susceptibility times a squared electric field term. Moreover, based on the Boltzmann kinetic formalism, the magnitude of the susceptibility tensor is proportional to $e^3 \tau D_{x (y)} / 2 \hbar^2  (1+ i  \omega \tau)$ \cite{battilomo2021anomalous,sodemann2015quantum}. So, the results are valid for an oscillating electric field caused by THz or microwave radiation types, where $\omega \tau \approx 1$. For a deduced $D_y=-10^{-10}$ m caused by $B_x= 8$ T and $B_y=1$ T (presented in Fig.~\ref{B}), in bilayer graphene under a planar magnetic field and considering $\tau= 0.15 \times 10^{-12}$ s, we can show that for $|E|=10^6$ V/m and $\omega=2.1 \times 10^{13}$ rad/s, the magnitude of the current density in bilayer graphene is $0.11$ A/cm. 
It has to be mentioned that the linear term of conductivity leads to a larger current density compared 
to the quantum nonlinear term, which we discussed here. However, the nonlinear term represents the topological aspect of the model that is to be considered as the corrections on the linear term.

The study of higher frequencies can be done by modifications on Boltzmann equation via quantum kinetic theory \cite{sodemann2019, xiao2019,asgari2022}, which could be considered in future works.
Additionally, at zero magnetic field each energy band has a degeneracy for spin--up and spin--down electrons. When a magnetic field is applied to the bilayer graphene, the degeneracy of spin--up and spin--down electrons is broken by the applied magnetic field and the energy difference between spin--up and spin--down electrons is a Zeeman--energy equal to $\Delta E_z= 2 S \mu_B B$, where $S$ is the spin of an electron and $\mu_B$ is the Bohr magneton. Considering $S=1/2$ leads to a $\Delta E_z=5.8 \times 10^{-5}$ eV/T. Hence, for a $2$ T magnetic field, we get $\Delta E_z \approx 10^{-1}$ meV. Besides, in our study, the Fermi level energy changes so that $0 \leq \mu \leq 40$ meV. According to the value of Fermi energy, we could expect
the total current density produced by spin--up and spin--down electrons could increase up to two times of calculated current density. 

\section{Conclusion}\label{conclusion}
This article represents an analytical study confirmed with numerical results of QNLPHE in bilayer graphene, which can be controlled by an in-plane steady magnetic field in the absence of spin--orbit coupling. The proposed strategies of a tunable BCD could also be applied to a wide range of other two--dimensional materials, such as phosphorene \cite{kheirabadi2021current}, which declares that
our findings pave the way to discover exotic nonlinear phenomena in 2D materials. We reveal a QNLPHE with a Hall--voltage that is quadratic with respect to the applied electric field.
The orbital--induced Berry dipole is strongly enhanced in AB--stacked bilayer graphene and reaches the nanometer scale. The aim of this work is to show that this topological effect emerges even in the complete absence of spin--orbit coupling in 2D Dirac materials, where two or more bands cross or nearly cross. Additionally, recently recognized optoelectronic and nonlinear transport experiments can give straight access to the dipole moment of the Berry curvature in non-magnetic and non-centrosymmetric materials \cite{battilomo2019berry, xu2018electrically, moulsdale2020engineering}. The predicted effects could also be utilized in applications that demand second--harmonic generation or rectification, which are used, for instance in wireless communications, infra-red detectors and energy harvesting applications. 
Moreover, such a magnetically switchable BCD may ease the observation of a broad range of quantum geometrical phenomena like the geometric properties of Bloch states in a large number of 2D materials and help consideration of other quantum geometrical phenomena \cite{battilomo2019berry, you2018berry}, and the facilitation of fabrication and up--scaling of the approach could allow exotic phases of matter attractive in twistronics \cite{ho2021hall}.

\section{Acknowledgement}
We would like to appreciate E. McCann, C. Ortix and R.  Asgari for fruitful discussions and useful comments.
We thank the Office of Vice President for Research of Sharif University of Technology for financial support.

\appendix
\section{Symmetry analysis}\label{A}
As mentioned in the main text, Eqs.~\ref{dx} and \ref{dy} are valid whenever $U_1=U_2$, which means that the underlying system has the spatial inversion symmetry. In this condition, the Berry curvature (Eq.~\ref{BC}) should also obey spatial inversion symmetry of the system, hence. Moreover, the honeycomb lattice is invariant under reflections with respect to x-- or y--axis. In this situation, we can show that the following identities are valid
\begin{eqnarray}\label{A1}
a_{m, n}(k_x,k_y) {B_x}^m {B_y}^n&&=a_{m,n}(k_x,-k_y) {(-B_x)}^m {B_y}^n\nonumber\\
&&=a_{m,n}(-k_x,k_y) {B_x}^m {(-B_y)}^n\nonumber\\
&&=a_{m,n}(-k_x,-k_y) {(-B_x)}^m {(-B_y)}^n.\nonumber\\
\end{eqnarray}
In the above equations, we have considered that under reflection with respect to x--axis $k_x \rightarrow k_x$,  $k_y \rightarrow -k_y$, $B_x \rightarrow -B_x$ and $B_y \rightarrow B_y$. Similarly, under the reflection with respect to y--axis, we have $k_x \rightarrow -k_x$,  $k_y\rightarrow k_y$, $B_x \rightarrow B_x$ and $B_y \rightarrow -B_y$. In addition, 
the inversion symmetry is given by $k_x \rightarrow -k_x$,  $k_y\rightarrow -k_y$, $B_x \rightarrow -B_x$ and $B_y \rightarrow -B_y$.

According to the definition of BCD (Eq.~\ref{bcd}) and the expression for Berry curvature proposed in Eq.~\ref{BC}, we will show that in the final expression of BCD either $m$ or $n$ must be odd. Those terms, which contain both odd or both even exponents ($m, n$) vanish in the final expression of BCD.
To prove this, we would like to stress that BCD comes out of an integration on Brillouin zone (BZ) which can be considered symmetrically around the origin of ${\bf k}$-space. The integrand of Eq.~\ref{bcd} for $D_x$ contains the following derivative,
\begin{equation}\label{3}
\frac{\Delta \Omega}{\Delta k_x}=\frac{\Omega(k_x+\Delta k_x, k_y, B_x, B_y)-\Omega(k_x, k_y, B_x, B_y)}{\Delta k_x}. \\
\end{equation}
According to the reflection symmetry with respect to $y$-axis, for each occupied state with $k_x>0$ there is a state at $-k_x$ in the BZ, which gives
\begin{equation}\label{4}
\frac{\Delta \Omega}{\Delta k_x}=\frac{\Omega(-k_x, k_y, B_x, B_y)-\Omega(-(k_x+\Delta k_x), k_y, B_x, B_y)}{\Delta k_x}.\\
\end{equation}
We rewrite the derivatives given in Eqs.~(\ref{3}, \ref{4}) using the expression for BC (Eq.~\ref{BC})
\begin{widetext}
\begin{equation}\label{5}
\frac{\Delta \Omega}{\Delta k_x}=\frac{\sum a_{m,n}(k_x+\Delta k_x, k_y) {B_x}^m {B_y}^n-\sum a_{m,n}(k_x, k_y) {B_x}^m {B_y}^n}{\Delta k_x},
\end{equation}
\begin{equation}\label{6}
\frac{\Delta \Omega}{\Delta k_x}=\frac{\sum a_{m,n}(-k_x, k_y) {B_x}^m {B_y}^n-\sum a_{m,n}(-(k_x+\Delta k_x), k_y) {B_x}^m {B_y}^n}{\Delta k_x}.
\end{equation}
\end{widetext}
In accordance with Eq.~\ref{A1}, when $m$ and $n$ are both even and odd numbers, we have
\begin{equation}\label{A2}
a_{m,n}(k_x,k_y)=a_{m,n}(-k_x,k_y).
\end{equation}  
Therefore, the corresponding terms (of both even and odd $m, n$) vanish in summing up the integration to obtain $D_x$. A similar explanation rule out the presence of both even and odd exponents in the final expression of $D_y$. That means, either $m$ or $n$ should be an odd integer.

In the next step, we consider the contribution of those terms with odd $m$ and even/zero $n$. Thus, we can assume that $m=2k+1$ and $n=2k'$ ($k, k'=0,1,2,..$), which leads to the following forms for Eqs.~\ref{3} and \ref{4},
\begin{widetext}
\begin{equation}\label{7}
\frac{\Delta \Omega}{\Delta k_x}=\frac{\sum a_{2k+1,2k'}(k_x+\Delta k_x, k_y) {B_x}^{2k+1} {B_y}^{2k'}-\sum a_{2k+1,2k'}(k_x, k_y) {B_x}^{2k+1} {B_y}^{2k'}}{\Delta k_x},
\end{equation}
\begin{equation}\label{8}
\frac{\Delta \Omega}{\Delta k_x}=\frac{\sum a_{2k+1,2k'}(-k_x, k_y) {B_x}^{2k+1} {B_y}^{2k'}-\sum a_{2k+1,2k'}(-(k_x+\Delta k_x), k_y) {B_x}^{2k+1} {B_y}^{2k'}}{\Delta k_x}.
\end{equation}
\end{widetext}   
The application of Eq.~\ref{A1} for odd $m$  and even $n$ gives the subsequent identity,
\begin{equation}
a_{2k+1,2k'}(k_x,k_y)=a_{2k+1,2k'}(-k_x,k_y).
\end{equation}
Consequently, the summation of Eqs. \ref{7} and \ref{8} in the integrand of $D_x$ vanishes. The only remaing terms, which lead to non-zero $D_x$ is even $m$ and odd $n$ as presented in Eq.~(\ref{dx}).
Similar arguments verifies the expression for $D_y$ presented in Eq.~(\ref{dy}), which completes our proof.
\begin{figure}[h]
   \centering   
   \includegraphics[scale=0.51]{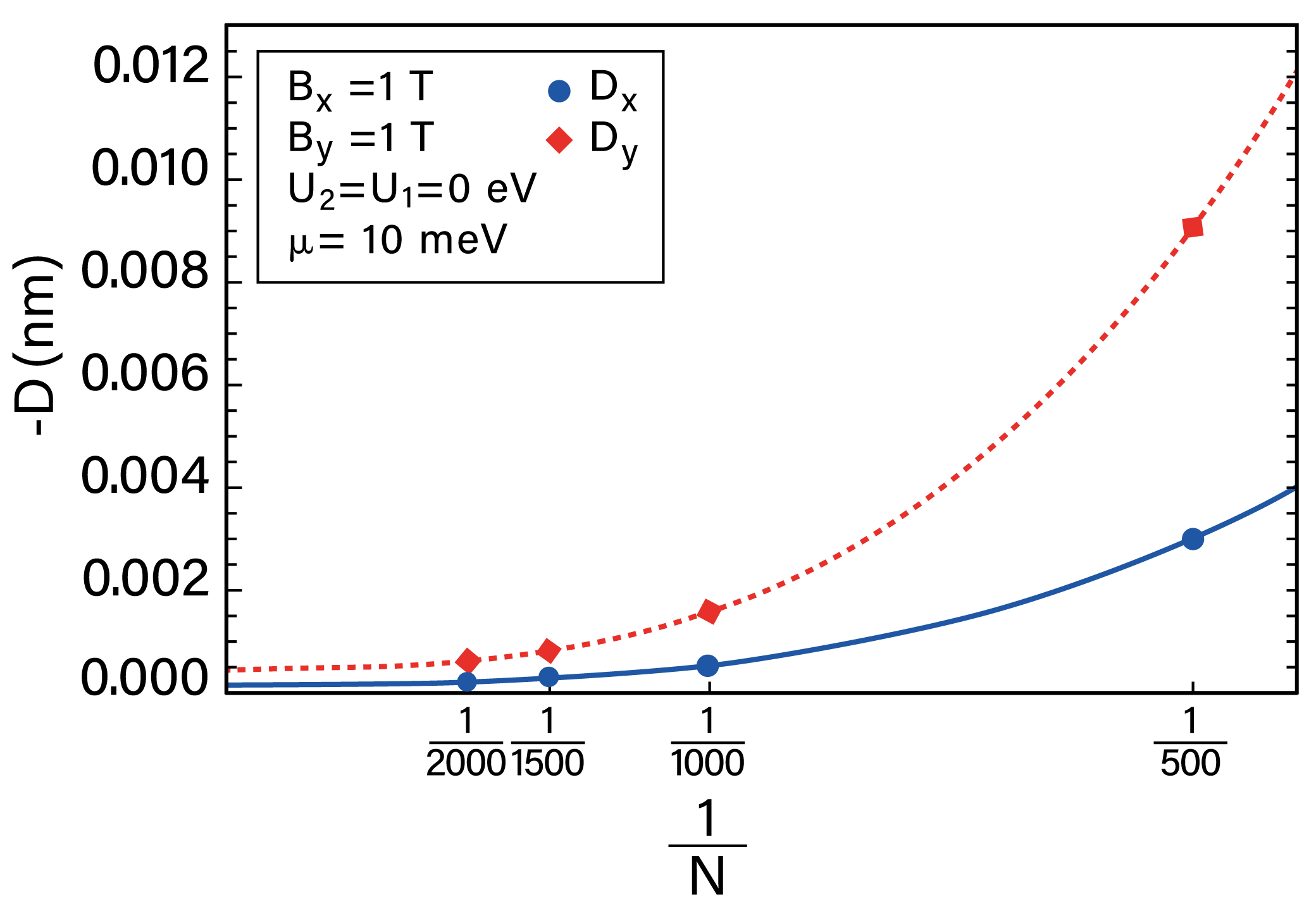}
   \caption{The dependence of $D_x$ and $D_y$ versus $\frac{1}{N}$, where $N$ is the number of mesh grids in each directions.  Solid line (dashed line) represents $D_x$ ($D_y$), where the other parameters are $B_x=B_y=1$ T, $\mu=10$ meV, $\Delta=0$ eV.} 
    \label{mesh}
\end{figure}
\section{Numerical method to calculate BCD}\label{BB}
The following numerical approach has been used to obtain BCD. 
Firstly, the BZ is split to discretized mesh of $k_x(i), k_y(j)$, where $k_x(i)=\frac{2 \pi i}{N}$ for $i=0, 1, \dots, N-1$ and similar values for $k_y(j)$.
Then, based on Eq.~\ref{omegan}, the discretized value of $\Omega_z^n(k_x(i), k_y(j))$ is assigned to a matrix of $N \times N$. It means that $\Omega$ is represented by a matrix in ${\bf k}$-space.
The derivative of the Berry curvature in the $x$ or $y$ direction is derived using a finite difference method. We implemented this approach at fixed parameters of the model for different mesh grids, i.e. $N$. Our results show that a convergence is obtained for $N \sim 2000$. 
We have calculated both $D_x$ and $D_y$ of our model at $B_x=B_y=1$ T, $\mu=10$ meV and in the absence of any gate voltage for different mesh grids as shown in Fig~\ref{mesh}. The horizontal axis is the inverse of mesh numbers in each directions, i.e. $1/N$. However, producing data with $N=2000$ requires high CPU time, for instance, to obtain a single point of BCD the CPU time of 14 days was spent on a machine with 80 cores. This leads us to stick on the value of $N=500$ to produce all data points and investigating the dependence of BCD on different parameters.
Although the BCD value changes drastically from $N=500$ to $N=2000$ the extrapolation to $N\rightarrow \infty$ gives nonzero result for BCD. Moreover, if we only keep the values of N=1000, 1500, 2000 we observe weaker mesh finite size effect, which justifies that final result would be non-zero.

\bibliography{bib} 

\end{document}